\documentclass[superscriptaddress,
				a4paper,
				aip,rsi, nofootinbib,
				amsmath,amssymb,
				reprint
				]{revtex4-1}
\usepackage[utf8]{inputenc}
\usepackage{graphicx}
\usepackage{amsmath,amssymb}
\usepackage{textcomp}
\usepackage{color}
\usepackage{xcolor}
\usepackage{epstopdf}
\usepackage{siunitx}
\usepackage[unicode=true, colorlinks=true, linkcolor = blue, citecolor = blue, urlcolor=blue]{hyperref}
\usepackage{mathtools}
\usepackage{natbib}
\usepackage{multirow}

\usepackage[normalem]{ulem}
\usepackage{soul}

\newcommand{\vect}[1]{\boldsymbol{#1}}

\begin{document}

\title{Two-axis goniometer for single-crystal nuclear magnetic resonance measurements}
\author{Ton\v ci Cvitanić}
\affiliation{Department of Physics, Faculty of Science, University of Zagreb, Bijeni\v {c}ka c. 32, Zagreb HR 10000, Croatia}
\author{Marin Lukas}
\affiliation{InTenX Solutions, St. Lucas d.o.o., Jabukovac 1, Zagreb HR 10000, Croatia}
\author{Mihael S. Grbić}
\email{mgrbic@phy.hr}
\affiliation{Department of Physics, Faculty of Science, University of Zagreb, Bijeni\v {c}ka c. 32, Zagreb HR 10000, Croatia}

\begin{abstract}
We report the design and construction of a two-axis goniometer capable of any sample orientation with respect to the external magnetic field. The advantage of this design is that it allows free rotations around a single axis independent of the other which minimizes rotational
error without reduction of angle range. Goniometer is capable of operating with high precision at both low and high temperatures and in high magnetic fields. It was mounted on the custom made nuclear magnetic resonance probe for use in Oxford Instruments wide-bore variable field superconducting magnet.
\end{abstract}

\maketitle

\section{Introduction}
In the field of solid-state physics it is not uncommon to work with materials of anisotropic crystal structure. Plenty of physical properties show anisotropies, such as transport, mechanical and optical properties, which all depend on the crystal orientation of external excitations. Magnetism brings even more anisotropies, as interactions can be anisotropic, e.g. single-ion anisotropy, Dzyaloshinskii-Moriya, staggered $g$-tensor component... Because of this vast number of reasons for material to posses anisotropy, there is a need for precise sample orientation during physical measurements. This allows us to control the properties of the system and determine all the anisotropic components or reduce the anisotropy by choosing a highly symmetric orientation.

Nuclear magnetic resonance (NMR) and nuclear quadrupole resonance (NQR) techniques are very powerful probes for investigation of local electric and magnetic fields that provide microscopic insight into chemical, electronic and spin structure of novel and interesting materials. Both techniques detect splitting of nuclear spin energy levels due to magnetic and electric interactions -- NMR refers to the case when splitting is mostly due to external magnetic field and NQR when it is due to electric field gradient (EFG) on nuclei site. In both cases, energy levels are described by a Hamiltonian:
\begin{eqnarray}
\mathcal{H} = -\gamma_n \hbar \mathbf{B}(\vartheta_B, &&\varphi) \mathbf{I} +\\ \nonumber
&&+ \frac{e Q \ V_{zz}}{4 I (2I - 1)} \left( 3 I_z^2 - I^2 + \eta (I_x^2 - I_y^2) \right),
\label{eq:ham}
\end{eqnarray}
where $\gamma_n$ is gyromagnetic ratio, $\mathbf{I}$ is nuclear spin, $eQ$ is quadrupolar moment of nucleus, $V_{zz}$ and $\eta$ define EFG tensor in its pricipal axes. Tensor component $V_{zz} = \partial^2 V / \partial z^2$ is the largest value and $\eta = (V_{yy} - V_{xx})/V_{zz} >0$ is the asymmetry parameter. Polar angle $\vartheta_B$ and azimuthal angle $\varphi$ describe the position of the external magnetic field $\vect B$ in the coordinate system of the EFG principal axes. Since these are microscopic techniques, they can detect crystal misalignments the precision of 1\textdegree\ in-situ.\\
\indent To properly determine interactions and their anisotropies of a specific material, a single crystal sample is required. Furthermore, for a local non-cubic symmetry at a specific atom site the EFG is nonzero which provides measurement anisotropy that one must take into account while performing NMR measurements. This anisotropy can occur even in the case of a cubic crystal cell. For these reasons, it is highly useful to include a two-axis goniometer in the NMR setup.\\
\indent The samples used in NMR investigations of condensed matter physics are usually small. To optimize the signal strength of the the nuclei of interest it is important for the measurement coil to be wound directly around the sample. This increases the filling factor \cite{Poole, Hoult_1976} (ratio of sample volume to coil volume: $\eta = V_s / 2 V_c$) and considerably reduces the signal losses. This imposes a restriction to the goniometer design: it is highly desired that goniometer has enough space to rotate the coil together with the sample. This excludes designs where the rotator is small so that it can be placed inside the (much larger) coil.\cite{Vosegaard_1996}\\
\indent Previous goniometer designs usually have limited available angle of rotation \cite{Shiroka_2012}, or include differential gear setup \cite{Herzog, Suzuki_1997}, where sample rotates around one rotational axis by turning both driving gears together in-phase, while rotation around other axis is achieved by rotating them in antiphase. Thus we can say that rotations are coupled together with some arithmetic relation connecting the angles of rotation. This can complicate precise rotation around a single axis and might increase error due to rotation of both driving gears. Other goniometers, that had decoupled rotations were constrained in terms of obtainable angles. Our approach was to have one axis freely rotating independent of the other to minimize rotational error while maintaining all possible obtainable angles.\\
\indent Below is a description of our two-axis goniometer setup. Its main novelties include high precision achieved with meticulous machining of small parts, functioning at high magnetic fields and in wide temperature range as it is made from non-magnetic materials of similar temperature expansion coefficient. At the same time, by using commercially available parts when available, low cost of manufacturing is achieved.

\section{Design and manufacture}
The goniometer was designed and manufactured to be used in a wide bore ($D = 60$~mm) variable temperature insert (VTI) of Oxford Instruments variable field superconducting magnet (up to $12$~T). For this it had to be mounted on a suitable probe that also contains NMR circuitry -- tuning and matching capacitors, soldering point for NMR coil and a semi-rigid coaxial cable that connects the NMR circuitry at one end to the top of the probe which leads outside the VTI. The probe was custom made to accommodate the goniometer and two turning rods to rotate the sample. To aid in rotations, two mechanical counters were added to the turning rods.

Goniometer was contrived with several important requirements: (1) to be precise in both axes; (2) to work in high magnetic fields and wide temperature range; (3) to be able to achieve any sample orientation; (4) to maximize sample space;

To meet the first two requirements the authors designed the goniometer with tight fitting of parts and by choosing dimensions that minimize backlash. This provides a precise rotation control. Since most of the parts fit together with another part of the same material, they can contract or elongate by a same amount, remaining in perfect fit at any temperature. Brass was chosen as the material for the manufacturing, since it is easily available and has good physical properties: it is non-magnetic, good at heat conduction (important for sample thermalization), rigid and easily machinable. Brass has a fairly good bearing properties, especially in the regime of non continuous rotations. Because of this, low-friction materials were not used for bearings, further reducing the risk of jam due to different materials in tight fit. Last two requirements were met by carefully designing parts. 

\begin{figure}[ht!]
\centering
\includegraphics[width=0.5\linewidth]{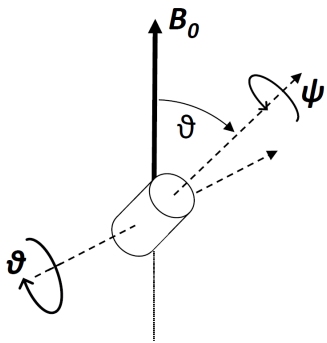}
\caption{Rotator geometry. $\vec B_0$ is external magnetic field which defines orientation of crystal. Angle $\vartheta$ controls azimuthal rotation while angle $\psi$ defines orientation in plane defined by azimuthal rotation. Two angles $\vartheta$ and $\psi$ are standard Euler angles.}
\label{fig:geometry}
\end{figure}

\begin{figure}[b!]
\centering
\includegraphics[width=0.9\linewidth]{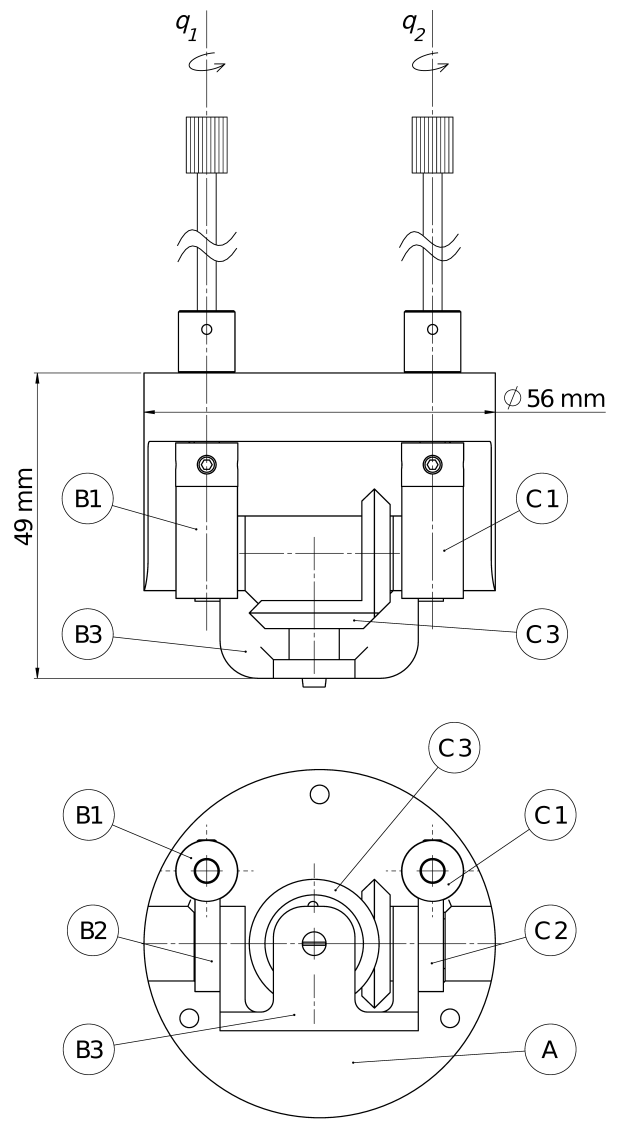}
\caption{Drawing of the two axis goniometer. \textbf{A}: the body of the goniometer; \textbf{B1-2} Worm and worm gear of $q_1$ rotation; \textbf{B3} The cradle that is driven by $q_1$ rotation; \textbf{C1-2} Worm and worm gear of $q_2$ rotation \textbf{C3} Bevel gear with sample mount (not shown for clarity) is directly driven by $q_2$ rotation through bevel and indirectly by $q_1$ by rotation of the cradle.}
\label{fig:drawing}
\end{figure}

The goniometer design is shown in Fig.~\ref{fig:drawing}. Bevel gear with sample mount (C3) is mounted on cradle (B3) that makes rotations of angle $\vartheta$ (Fig.~\ref{fig:geometry}). Other rotation (of angle $\psi$ in Fig.~\ref{fig:geometry}) is made by rotation of the aforementioned bevel (C3). Angles $\vartheta$ and $\psi$ are standard Euler angles: $\vartheta$ is azimuthal rotation and $\psi$ is rotation around vector defined by azimuthal rotation. Mechanical rotations are carried with two worm gears with gear ratio 36:1. One worm wheel (B1) is directly connected to cradle and the other (C1) is connected to axle that freely rotates around cradle. Axle rotates bevel gear (C3) which turns sample.
The sample is mounted on an exchangeable sample mount (not shown) that is fastened to bevel gear C3. Mount should be made from rigid insulating material (such as POM, PIA, PEEK \ldots), and its height places the sample at the crossing of rotation axes. 
 
The two bevel gears are identical and have an external diameter of 20.7 mm with a flat surface on top, which is 14.8 mm in diameter. This gives a large sample space and provides an easy mount of electrical connections.

Rotation of the sample is done as follows: for the $\vartheta$ rotation, both worm gears must be turned by the same amount. For the $\psi$ rotation, only the worm (C1) is turned. If we define $q_1$ and $q_2$ to be input rotations made to the worms C1 and B1 respectively, we can write the angle of rotations as:
\begin{align}
\vartheta &= \frac{q_1}{36} \ , \\
\psi &= \frac{q_2 - q_1}{36} \ .
\end{align}
The numerator value (36) comes from the gear ratio of 36:1, i.e. if one makes a full turn with the worm ($q_i = 360$\textdegree) worm gear rotates 10\textdegree. Standard goniometer orientation with $(\vartheta, \psi) =$ (0\textdegree,0\textdegree) is as shown on the figure \ref{fig:drawing}. From this orientation, the cradle ($\vartheta$) can rotate $\pm 90$\textdegree, while $\psi$ rotation is constrained only by the physical contacts of the NMR coil. By carefully choosing lead lengths for the coil, a full rotation ($\pm 90$\textdegree) is easily obtainable. These two rotations provide every possible orientation a sample can have. During the operation care should be taken to always approach the desired orientation from the same side (e.g. by clockwise rotation of both worms), as to reduce backlash.

Worm gears and bevel gears were acquired commercially. We used a commercial machine shop to manufacture support structures (body and cradle) and modify the gears to our design. Axles and gear modifications were made with high-precision wire-EDM machine. Manual polishing was used where extra-tight fit was required. 

\section{Applications and discussion}
\begin{figure}[ht!]
	\centering
	\includegraphics[width=0.75\linewidth]{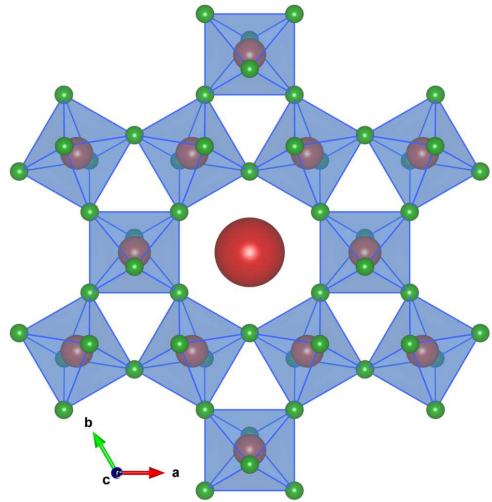}
	\caption{Structure of a single kagome pattern of Cs$_2$Cu$_3$SnF$_{12}$, as seen along the crystal $c$ axis. Copper, cesium and fluorine are colored orange, red and green, respectively. Tin is not shown for clarity. The structure shows CuF$_6$ octahedra tilted with "all-in"/"all-out" pattern.}
	\label{fig:struct}
\end{figure}

\begin{figure}[h!]
	\centering
	\includegraphics[width=1\linewidth]{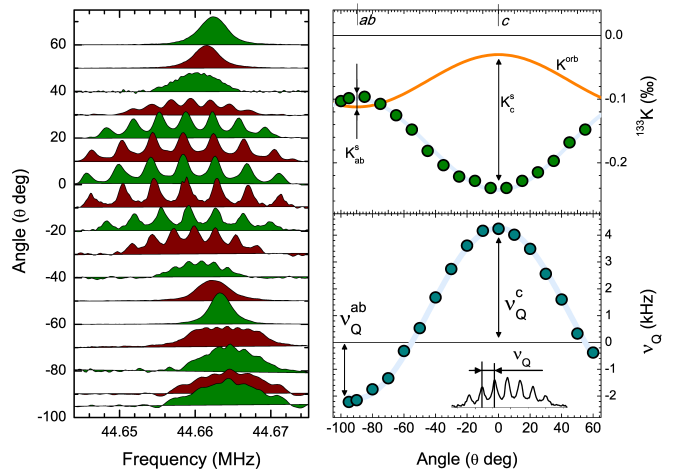}
	\caption{$^{133}$Cs NMR results at magnetic field $B_0 = 8$~T. Left: rotation spectrum around axis parallel to the kagome plane. Line shift and quadrupolar splitting change with angle is visible. Upper right: frequency shift calculated from the central line position. Full orange line is orbital part extrapolated from Clogston -- Jaccarino plot (not shown). Note the very small size of the shift (on the order of $10^{-4}$). Lower right: quadrupolar splitting as the average line split (denoted in the inset), and the small quadrupolar splitting (because of the small $^{133}$Cs quadrupolar moment). Since $V_{zz}$ of $^{133}$Cs is along the $c$-axis $\vartheta_B = \vartheta$.}
	\label{fig:CsKW}
\end{figure}

The goniometer was tested by $^{133}$Cs NMR measurement of the Cs$_2$Cu$_3$SnF$_{12}$ compound. The system consists of kagome planes that are formed by Cu$^{2+}$ ions connected via Cu-F-Cu bonds shown in Fig.~\ref{fig:struct}. Kagome planes extend in crystallographic $ab$ planes. From the structure of the compound~\cite{Tanaka} it can be seen that the Cs site occupies Wyckoff position $6c$, which means that it has a $3 m$ local symmetry. This will result in an axial ($\eta = 0$) EFG with the principal axis along the $c$ axis of the crystal (perpendicular to the kagome planes). This makes $^{133}$Cs NMR perfect for testing the goniometer, as observation of its quadrupolar splitting is well understood~\cite{slichter,abragam} and we know what is expected for the case of $^{133}$Cs nucleus (spin I$=7/2$).
In Fig.~\ref{fig:CsKW} we show the angle dependence of the $^{133}$Cs spectra measured at magnetic field value of 8 T, when the axis of rotation is in the $ab$ plane. From the spectrum at each angle value we can extract two informations: the quadrupolar splitting value $\nu_Q$ (shown in the lower right panel) and the frequency shift of the total spectrum (shown in the upper right panel).As we can see, the ratio of the $\nu_Q$ values at 0\textdegree\ and 90\textdegree\ is $\nu_Q (0^\circ)/\nu_Q (90^\circ) = -2$ within the error bar of the measurement (less than the symbol size).
This is exactly what is expected for $\eta = 0$ case, since then by definition we have $V_{xx} = V_{yy}$, and from Laplace equation ($\nabla^2 V = \sum\limits_i V_{ii} = 0$) it follows $V_{zz}= -2 V_{xx}$.

The frequency shift of the NMR spectrum consists of an orbital ($K^{orb}$) and spin ($K^{s}$) contributions:
\begin{equation}
K_i (T)= K_i ^{orb} + K_i ^s (T),
\label{eq:knight}
\end{equation}
where $i$ marks the orientation of the applied field. Since the contribution to $K^{orb}$ comes from the closed orbitals, the only temperature dependence of $K$ comes from the spin part $K^s$. Just like before, from symmetry it follows that $K_a = K_b$. By plotting the $K-\chi$ dependence (not shown), i.e. the Clogston -- Jaccarino plot, and understanding that in the high-temperature paramagnetic limit $\chi \rightarrow 0$ and the value of $K=K^{orb}$, we can extract $K_{ab, c}^{orb}$ values. From the measured data, we can extract all the values of the NMR parameters, shown in Table~\ref{tbl:params}.
\begin{table}[ht!]
	\centering
	\caption{Table of NMR parameters. Orbital parts of Knight shift were extracted from comparison of Knight shift with susceptibility (not shown). Spin parts of Knight shift were deduced from measurements in figure \ref{fig:CsKW}, using equation \ref{eq:knight}.
	}
	\label{tbl:params}
	\begin{tabular}{lrc}
		\hline
		\hline
		parameter & & value:   \\
		\hline
		$K_c ^s$ (\textperthousand) & & $-0.213 \pm 0.002$  \\ 
		$K_{ab} ^s$(\textperthousand) & & $0.013 \pm 0.002$ \\
		
		$K_c^{orb}$(\textperthousand) & & $-0.036 \pm 0.004$  \\
		$K_{ab} ^{orb}$(\textperthousand) & & $-0.1125 \pm 0.0002$ \\
		$\nu_Q$ (kHz) & & $4.26 \pm 0.04$  \\
		$\eta$ & &0  \\
		\hline
		\hline		
	\end{tabular}
\end{table}
These values are extremely small for a typical research of strongly correlated systems, and can only be properly measured with a stable measurement setup. 

We also used the goniometer to determine the symmetry of the $^{63}$Cu site in the same compound. From the structural data it follows that the Cu site here occupies the $9e$ Wyckoff site and that it has $.2/m$ symmetry. However, on closer inspection, it can be seen that the local environment is a CuF$_6$ octahedron, where the two apical fluorine sites (F$^{apic.}$) are considerably farther than the four planar F ions: length of the Cu$-$F$^{apic.}$ bond is 2.348~\AA, while the length of the planar Cu$-$F bonds is 1.8997~\AA. This determines that the electron on the copper site resides in the $d_{x^2-y^2}$ orbital. Since the distances between the four planar Cu$-$F bonds are all equal and only the angle of the nearest neighbor F$-$Cu$-$F bonds varies, we can expect that the observed EFG symmetry of $^{63}$Cu will deviate from tetragonal bellow the sensitivity of the NMR measurement. This claim is also confirmed by a point-charge calculation of the EFG tensor that will be published elsewhere.  

The size of the quadrupolar splitting of $^{63}$Cu in this compound is large enough to be observed in zero field and an NQR line was found at 52.684 MHz at 190 K. If we would like to determine the local symmetry of the $^{63}$Cu site in the same way as we did with $^{133}$Cs due to such large quadrupolar splitting it would require a very large magnetic field. It is therefore much more convenient to apply a small external magnetic field and rotate the sample with respect to $\vect B_0$. The small Zeeman perturbation will split the NQR line~\cite{abragam,Kind_1986,Garcia_1986} and from its angle dependence we can determine the local symmetry of the site.

\begin{figure}[ht!]
	\centering
	\includegraphics[width=0.7\linewidth]{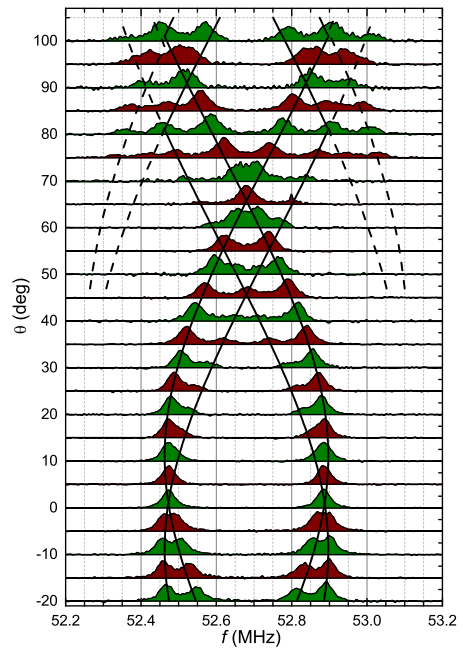}
	\caption{$^{63}$Cu NQR spectrum with a small external magnetic field $B_0 \approx 35$~mT. The sample was rotated along an axis inside the kagome planes. Spectra were taken at $T = 190$~K.}
	\label{fig:CuTETh}
\end{figure}

First we show the dependence on the angle $\vartheta$ in Fig.~\ref{fig:CuTETh}, with the applied magnetic field $B_0 = 35$ mT and the axis of rotation again in the $ab$ plane. The CuF$_4$ plaquettes are tilted with respect to the kagome planes at an angle of 12.6\textdegree. They are located in the triangles of the kagome lattice and create an "all-in"/"all-out" pattern. Therefore, when the magnetic field is oriented perpendicular to the kagome planes ($\vartheta = 0$\textdegree), we will observe one pair of lines, but as the angle changes we will observe two sets of signals - one where the initial mean tilting is 12.6\textdegree\ and one where the initial tilting is -12.6\textdegree. In reality there is a distribution of the tilting angles that is set by the orientation of the component of the magnetic field within the kagome plane. However, the fine details are obscured by the finite width of the NQR lines and only two sets of patterns are visible.      

The measured angle dependence of the spectrum is compared to the analytical solution for an axial $\eta=0$ case shown in Fig.~\ref{fig:CuTETh}, and it is visible that the measurements can be well described by this function. The dependence reproduces the behaviour of the main NQR lines (full line) as well as the so-called forbidden transitions (dashed). 

\begin{figure}[t!]
\centering
\includegraphics[width=0.99\linewidth]{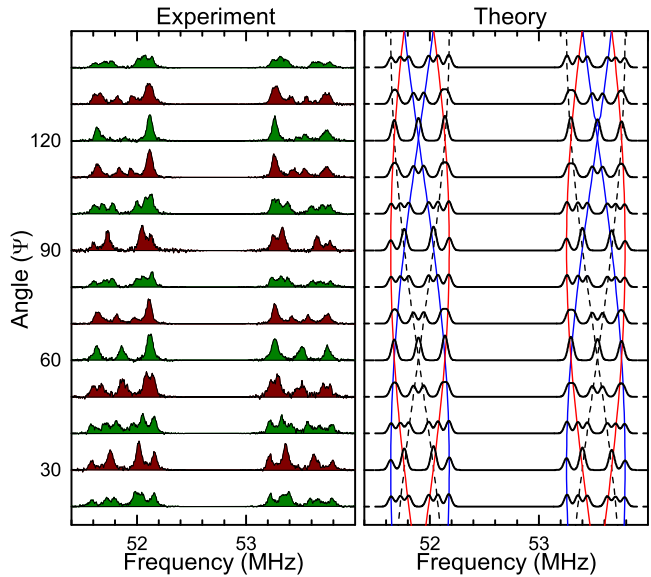}
\caption{$^{63}$Cu NQR with small external magnetic field $B_0 \approx 70$~mT. Spectra were taken at $T = 190$~K. Comparison of experiment and theory. Parameters for theoretical fit are in table \ref{tbl:params}. During the rotation, spectra follow three sinusoidal curves, each 120\textdegree\ apart, shown as red, blue and dashed. This shows the underlying crystal symmetry. The angle $\Psi$ lies within the $ab$ plane, as described in the text.}
\label{fig:CuTE}
\end{figure}

To additionally check the consistency of the data, we also measured the angle dependence of the NQR spectra for case when the small magnetic field is rotated within the kagome planes, i.e. the axis of rotation is along the crystal $c$ direction (left panel of Fig.~\ref{fig:CuTE}). To have a better resolution of the produced pattern, we have applied a somewhat larger field of $B_0 \approx 70$~mT. From the crystal structure we expect three pairs of lines that have identical angle dependence, but shifted in phase for 120\textdegree. The overlapping of lines will create a pattern visible in the right panel of Fig.~\ref{fig:CuTE}, which was generated by exact diagonalization of the NMR/NQR hamiltonian. In the figure the angle $\Psi$ moves in the $ab$ plane, but since the $V_{zz}$ of each copper site is tilted away from the $c$-axis, it does not match the angle $\varphi$ for $\vartheta_B=90$\textdegree, but rather it moves from ($\vartheta_B=90$\textdegree,$\phi =0$\textdegree) to ($\vartheta_B=77.4$\textdegree,$\phi =90$\textdegree), for $\Psi=0$\textdegree and $\Psi=90$\textdegree, respectively. 

It is visible that the simulated spectra nicely reproduce the measured data. It properly reproduces the distribution of the spectral intensity (in particular the high-symmetry points visible at e.g. 60\textdegree) but not fully the intensity of specific lines. However, our simulation does not take into account the change of intensity due to the change of the coil orientation with respect to $\vect B_0$, or the variations of the optimal pulse across the spectrum, which can explain the small variations in intensity.

The fitting of simulated spectra gives  $B_0~=~72.6$~mT, $\theta = 12.7$\textdegree\ and $\eta =0$, while $\nu_Q$ is set by the pure NQR value in zero field. The fitted parameters are in excellent agreement with what is expected from the crystal structure. Therefore, this demonstrates that our two-axis rotator behaves as expected and shows no measurable deviations, even when tested by a sensitive local probe. 

In conclusion, we have designed a two axis rotator where one of the axis of rotation is separated from the other which minimizes error during the measurements. The designed apparatus has been tested by a series of tests with NMR, a local probe that can detect even the smallest misalignments in-situ. The reported setup is suitable for other measurement techniques as well, not just NMR. There are no limitations on the number of wire connections, and only limitation is the sample space at the cradle of the goniometer which is sufficient.

\section{Acknowledgements}
Authors are grateful to Zlatko Kvočić, dipl. ing. meh. for his advice and input during the design process, and his time and patience in the manufacturing of small precise parts for the goniometer. Also, we would like to thank H.~Tanaka for providing the single crystal of Cs$_{2}$Cu$_{3}$SnF$_{12}$.

The authors acknowledge the support of the Croatian Science Foundation (HRZZ) under the project IP-2018-01-2970, and the Unity Through Knowledge Fund (UKF Grant No. 20/15).

%

\end{document}